\documentclass[12pt]{article}
\usepackage{graphicx}
\usepackage{amsmath}
\usepackage{amssymb}
\usepackage{booktabs}
\usepackage{url}
\usepackage[utf8]{inputenc}
\usepackage[T1]{fontenc}
\usepackage{hyperref}
\usepackage[a4paper, margin=2.5cm]{geometry}
\usepackage[numbers]{natbib}

\begin{document}

\title{Measuring Scientific Group Performance: Integrating h-Group and
Homogeneity into the $\alpha$-Index}
\author{ Roberto da Silva$^1$ \and Jos\'{e} Palazzo M. de Oliveira$^{2,3}$
\and Viviane Moreira$^{2,3}$ \\
%EndAName
$^{1}$Instituto de F\'{\i}sica, $^{2}$Instituto de Informatica, $^{3}$PPGC \\
Universidade Federal do Rio Grande do Sul (UFRGS) \\
Caixa Postal 15.064 -- 91.501-970 -- Porto Alegre -- RS -- Brazil \\
\texttt{rdasilva@if.ufrgs.br}, \texttt{\{palazzo, viviane\}@inf.ufrgs.br} }
\date{}
\maketitle

\begin{abstract}
Ranking groups of researchers is important in several contexts and can serve
many purposes, such as the fair distribution of grants based on the
scientist's publication output, the concession of research projects, the
classification of journal editorial boards and many other applications. In
this paper, we propose a method for measuring the performance of groups of
researchers. The proposed method is called \emph{the $\alpha$-index}, and it
is based on two factors: (i) the homogeneity of the \emph{h-indexes} of the
researchers in the group; and (ii) the h-group, which is an extension of the 
\emph{h-index} for groups. Our method integrates the concepts of homogeneity
and absolute value of the \emph{h-index} into a single measure, which is
appropriate for the evaluation of groups. We report on experiments that
assess computer science conferences based on the \emph{h-indexes} of the
members of the program committee.
\end{abstract}

\section*{Keywords}

Gini coefficient, \emph{h-index}, Metrics in Science, Bibliometrics.

\section{INTRODUCTION}

Ranking and classification of researchers is among the most discussed topics
in the academic community in the last decades \cite%
{hirsch2005,laherrere1998,redner1998}. Rankings are useful for the fair
distribution of grants to researchers according to their excellence, and can
also be used to classify journals according to the quality of their
editorial boards. In the scope of this paper, the term \emph{quality} refers
to the research output as measured by scientific publications.

Numerical data on the distribution of citations has been extensively
explored by the scientific community, and universal laws have been
established. Based on the ISI (Institute for Scientific Information)
database, \cite{laherrere1998} suggest that the number of papers with $x$
citations decays as a power law $N(x)\sim x^{-\alpha}$ where $\alpha \sim 3$%
. Similarly, \cite{redner1998} found a stretched exponential form $N(x)\sim 
\exp[-(x/x_0)^\beta]$ with $\beta \sim 0.3$ when analysing data from the
1120 most-cited physicists between 1981 and 1997. \cite{hirsch2005} reduced
the complexity of the data distribution to quantify the importance of a
scientist's research output into a single measure known as the \emph{h-index}%
. Despite being controversial, the \emph{h-index} is widely employed by many
research funding agencies and universities all over the world. Hirsh's
simple idea is that a publication is good as long as it is cited by other
authors, i.e. "a scientist has index $h$ if $h$ of his $N_p$ papers has at
least $h$ citations each. The other ($N_p - h$) papers have $\leq h$
citations each, with $0\leq h\leq N_p$. There are alternatives to the use of
the \emph{h-index}; \cite{braun2004}, for example, uses the total number of
citations to quantify research performance. However, in this paper, we have
opted to use the \emph{h-index} because it is less prone to being inflated
by a small number of big hits or by the eminence of co-authors. Since its
proposal, it has become widely accepted and has been employed as the basis
for many scientometrics and bibliometrics research.

The problem addressed here is how to characterise and classify a group of
researchers considering the \emph{h-indexes} of its individual components.
The method we propose assumes that quality cannot be characterized just by a
high average \emph{h-index} for the group, but also by its homogeneity. Our
rationale is that a group can have a high average \emph{h-index} just by
having one very productive researcher. However, a homogeneous group with an
equivalent \emph{h-index} will be better, as homogeneity denotes greater
robustness of the group.

In this paper, we introduce a new method to measure the scientific research
output of a group of researchers. The proposed method quantifies the quality
of a group using a parameter that we call \emph{$\alpha$-index}. The \emph{$%
\alpha$-index} of a group is based on two concepts:

\begin{itemize}
\item the \emph{h-group}, which is an extension of the \emph{h-index} for
groups. It is measured by taking the maximum number of researchers in the
group, satisfying \emph{h-index} $\geq$ \emph{h-group}. The remaining
researchers in the group have \emph{h-index} $<$ \emph{h-group}; and

\item a known statistic employed to demonstrate the social inequality of a
country, the Gini coefficient \cite{gini1921,gini_wiki}.
\end{itemize}

An important fact in the fields of computer science and engineering is that
not only publications in journals are valuable, but publications in
qualified conferences and workshops also play an important role \cite%
{zhuang2007}. Thus, the analysis of the quality of the conferences is
important to enable a suitable evaluation of the researcher's production. In
this paper, we show how our proposed \emph{$\alpha$-index} can be used to
evaluate the quality of a conference based on the \emph{h-indexes} of the
members of its Technical Program Committee (TPC).

Our method was designed to perform a fair comparison between groups with
different sizes, which is adequate for analysing conferences, as different
conferences will have TPCs of different sizes. The results of our proposed 
\emph{$\alpha$-index} are consistent, and the relative ordering of the
groups remains the same even if a subset or superset of the groups is
compared.

In our tests, we collected and compiled bibliometric data for seven
conferences. These data include the individual \emph{h-index} of each TPC
member and the number of citations for their papers.

The remainder of this paper is organised as follows: Section 2 describes the
conferences used in our tests and their classification according to CAPES 
\cite{capes2010}, a Brazilian research funding agency responsible for
evaluating the quality of university graduate programs. In this same
section, we describe some statistical properties of the collected data and
compare them to some expected results found in the literature. Section 3
shows how the Gini coefficient is a natural definition to measure the
homogeneity of the program committees of scientific conferences in an
analogy to the homogeneity of the wealth distribution in a population. We
also define the \emph{h-group} and the \emph{$\alpha$-index} of a group. In
section 4, the main results are presented. Finally, in section 5 the
conclusions are presented and extensions of the model are briefly discussed.

\section{Preliminaries and previous statistics about conferences}

CAPES \cite{capes2010} has defined a system for classifying the estimated
quality of publication venues. The system is called \emph{Qualis}, and it
grades venues into three categories: A, B, or C. According to this grading
scheme, A is the highest quality, and it is usually assigned to top
international conferences. The criteria analysed include the number of
editions of the conference and its acceptance rate.

Table \ref{Table:conferences} shows data collected for seven conferences for
the same validity period as the official ranking. The \emph{h-indexes} were
extracted using the free software "Publish or Perish"\footnote{%
\url{http://www.harzing.com/resources.htm}} which collects citation data
from the Google Scholar service\footnote{\url{http://scholar.google.com/}}.

\begin{table}[h]
\caption{Conferences used for this work and their classification. The second
column shows the average \emph{h-index} for the TPC with the associated
standard error $(var(h)/n)^{1/2}$}\centering
\begin{tabular}{lcc}
\toprule Conference & $\langle h \rangle \pm (var(h)/n)^{1/2}$ & \#TPC
members \\ 
\midrule Conf. A (A) & 12.78 $\pm$ 0.65 & 207 \\ 
Conf. B (C) & 11.92 $\pm$ 1.60 & 27 \\ 
Conf. C (B) & 11.63 $\pm$ 1.55 & 67 \\ 
Conf. D (A) & 10.10 $\pm$ 0.66 & 102 \\ 
Conf. E (B) & 08.07 $\pm$ 0.83 & 87 \\ 
Conf. F (A) & 07.94 $\pm$ 0.69 & 39 \\ 
Conf. G (C) & 07.56 $\pm$ 2.39 & 16 \\ 
\bottomrule &  & 
\end{tabular}%
\label{Table:conferences}
\end{table}

As a starting point, we explore some preliminary statistics about these
conferences. It is interesting to check similarities between the properties
obtained from the TPC population and the properties expected from the
general scientific population. The first analysis was to plot the number of
citations to papers written by the TPC members as a function of their \emph{%
h-indexes}. These plots are shown in Figure 1.

\begin{figure}[h]
\centering\includegraphics[width=0.35\textwidth]{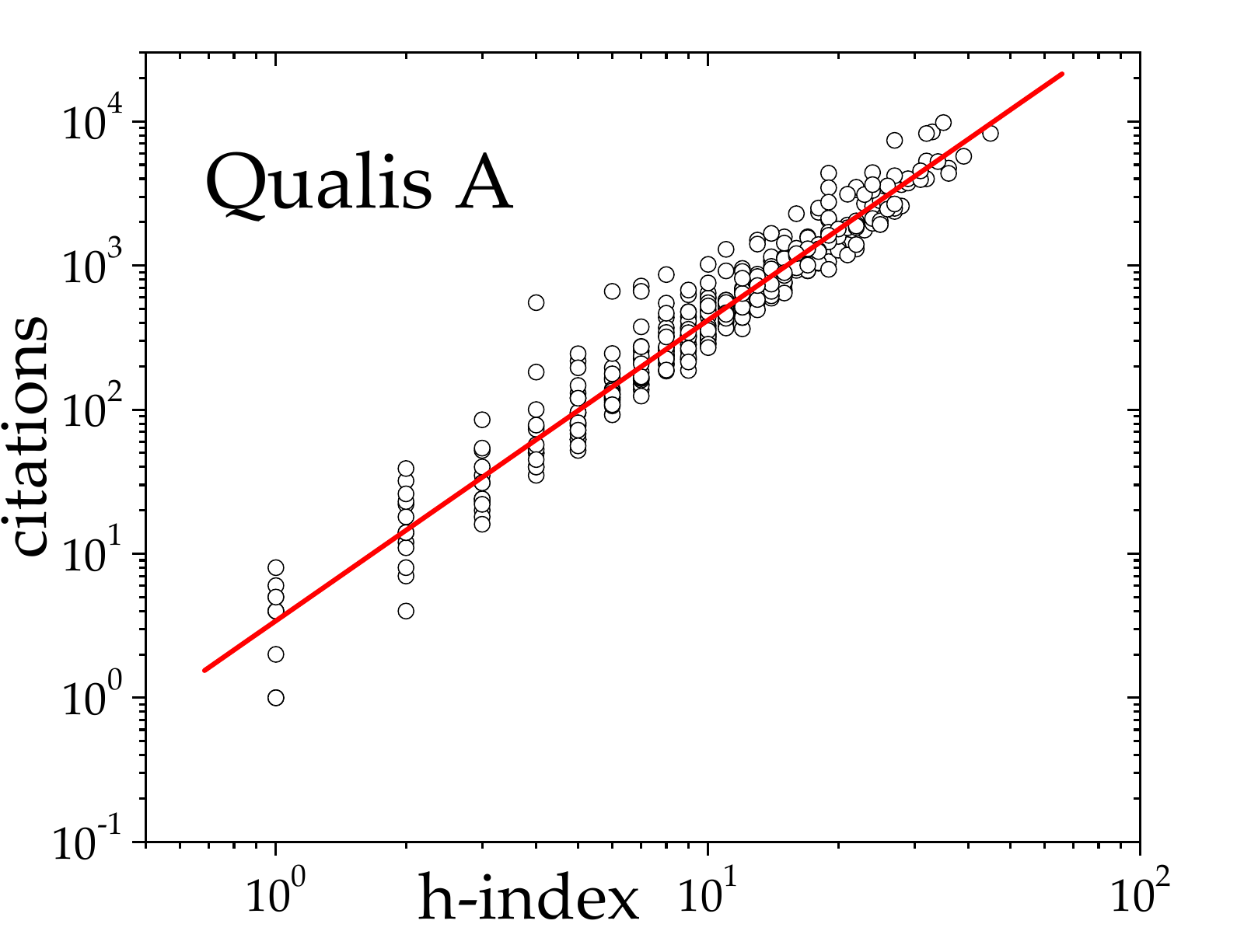}%
\includegraphics[width=0.35\textwidth]{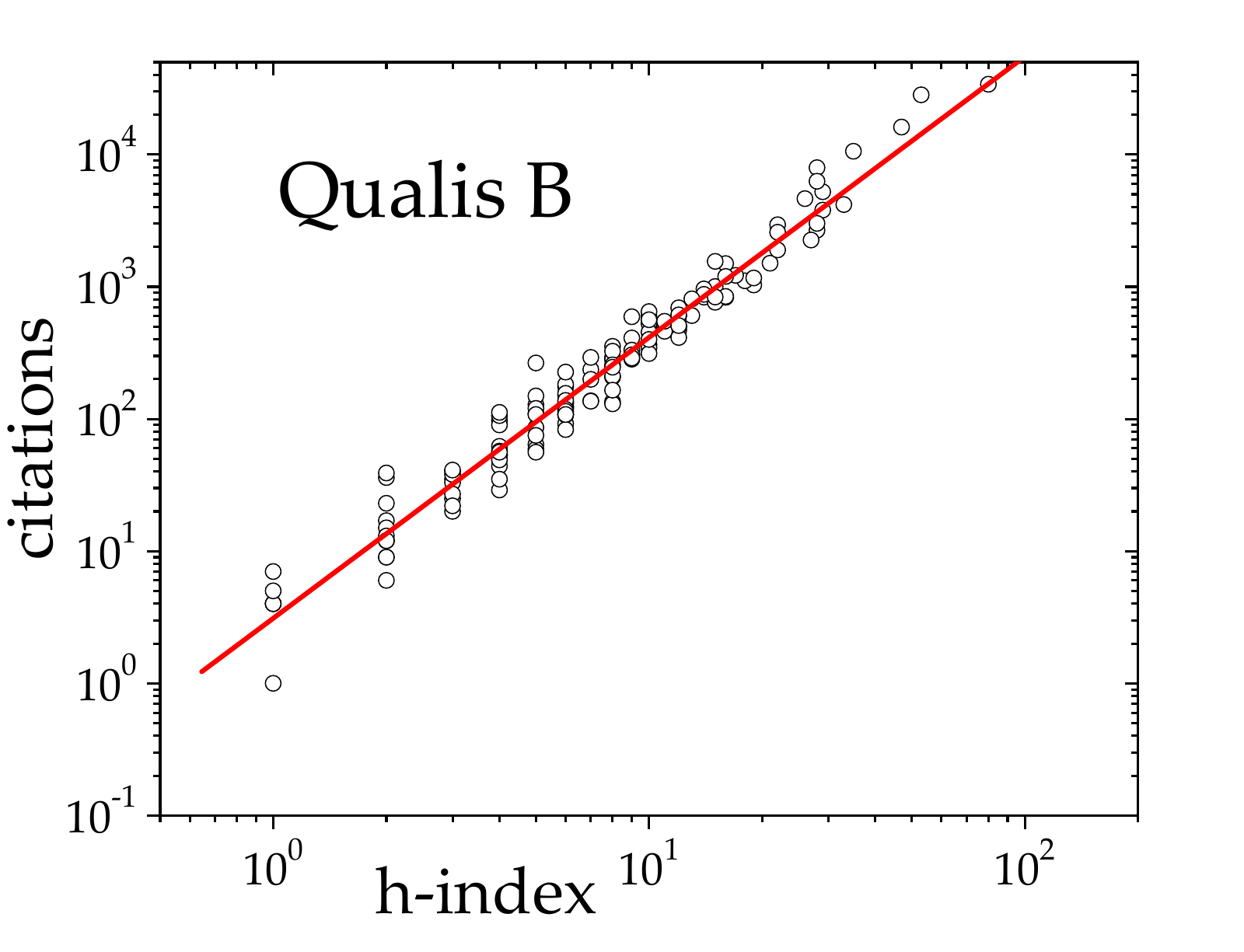}%
\includegraphics[width=0.35\textwidth]{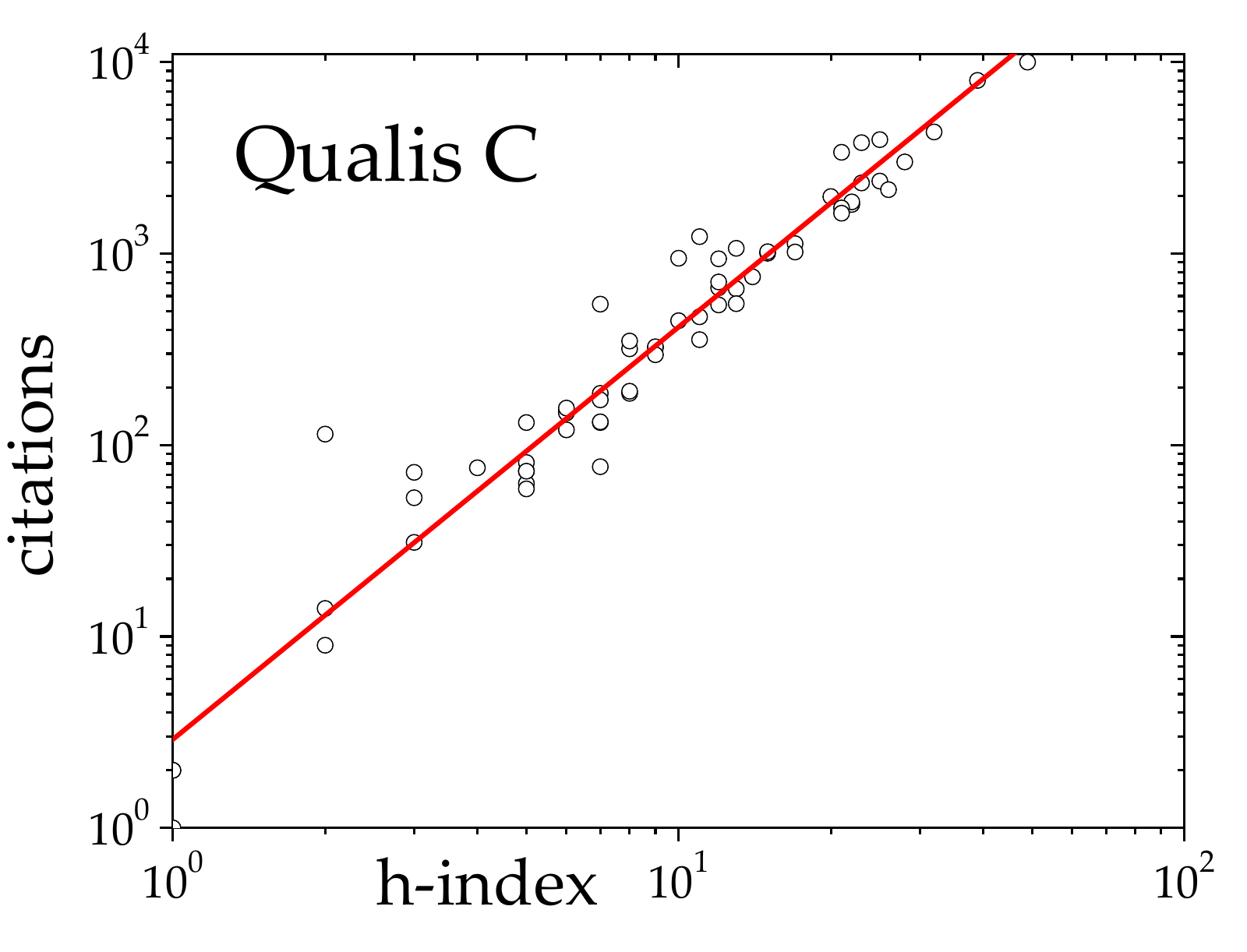}
\caption{Citations versus \emph{h-index} for conferences Qualis A, B, and C.}
\label{Fig:H_qualis_A_B_C}
\end{figure}

In all cases presented in Figure \ref{Fig:H_qualis_A_B_C}, we found that the
number of citations the authors have has a quadratic dependence on their
h-indexes, as obtained in scientific databases such as ISI (see, for
example, \cite{laherrere1998,redner1998}). In order to measure the $\alpha $
exponent, we separated our data according to the classification of the
conference (A, B or C) assigned by CAPES. For each set of conferences, we
analysed the expected relation $x\sim h^{\alpha }$, where $x$ is the number
of citations of the author and $h$ is the corresponding \emph{h-index}. In a
log-log plot, shown in Figure 1, we measured the slope. The results were $%
\alpha =2.08(3)$, 2.12(3), and 2.15(6) for conferences A, B, and C,
respectively. This result corroborates Hirsh's theory \cite{hirsch2005} in
which $\alpha =2$.

We analysed the distribution of citations in order to compare the features
of our data against the properties found in other scientific populations.
The analysis takes all TPC members into consideration (combining A, B, and C
conferences). The idea was to verify whether the distribution of the number
of citations, denoted by $x$, for TPC members of computer science
conferences follows a stretched exponential form:

\begin{equation}
N(x)\propto \exp \left[ -(x/x_{0})^{\beta }\right]  \label{eq:stretched_exp}
\end{equation}%
as claimed by Laherrere and Sornette in \cite{laherrere1998}. In their
study, they found $\beta \sim 0.3$, which can be determined by plotting a
histogram with the number of citations ($x$), as shown in Figure \ref%
{Fig:citations_and_ratio} (left plot).

\begin{figure}[h]
\centering\includegraphics[width=0.5\textwidth]{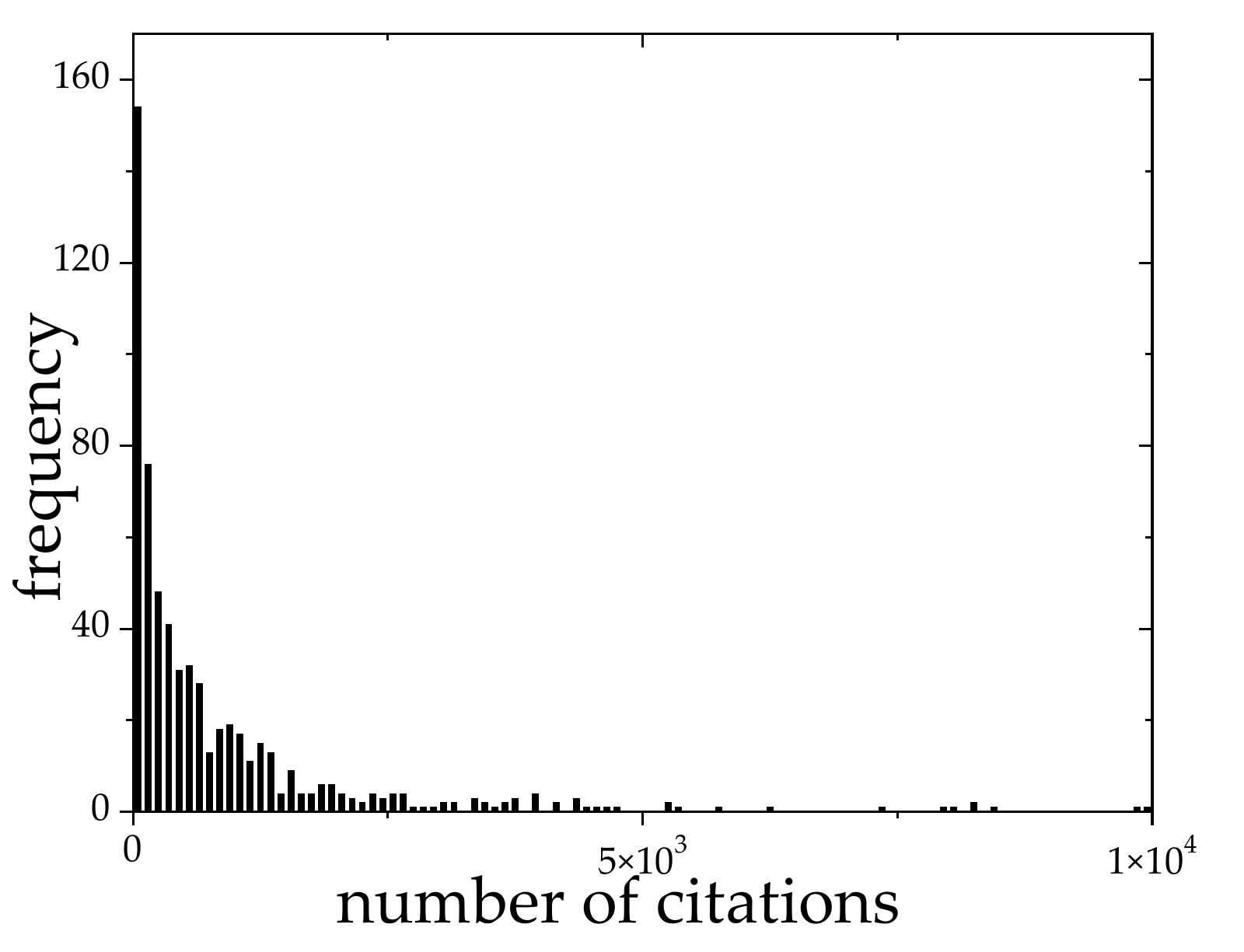}%
\includegraphics[width=0.5\textwidth]{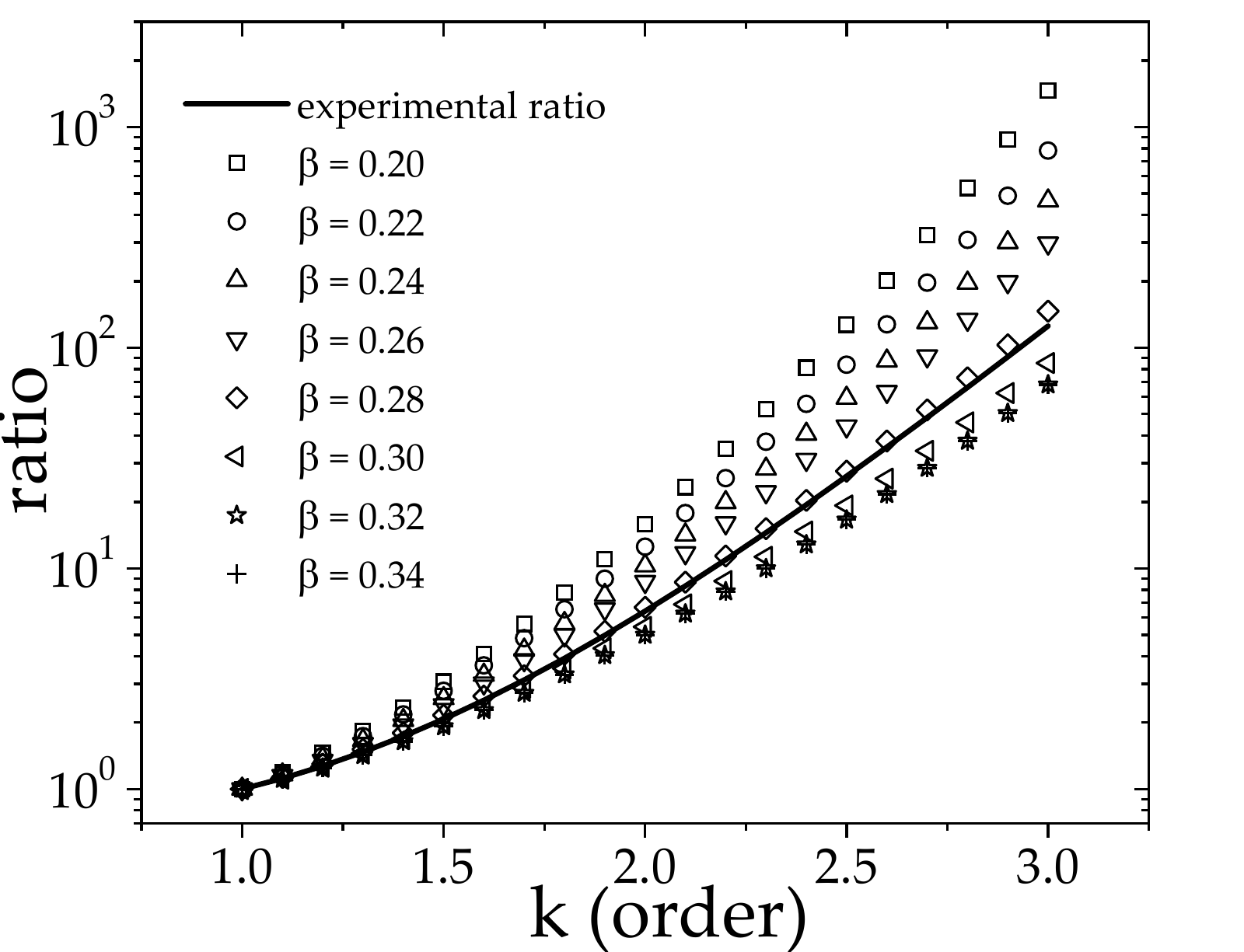}
\caption{(Left plot) Experimental distribution of the number of citations
for data obtained from conferences Qualis A, B, and C. (Right plot) A
comparison between the exact moments (Equation 2) obtained for different $%
\protect\beta $ values and the real moment, i.e. obtained from the collected
data (see Equation 3).}
\label{Fig:citations_and_ratio}
\end{figure}

Estimating $\beta $ directly from the data in Figure \ref%
{Fig:citations_and_ratio} (left plot) can be achieved by determining the
slope of the linear fit of $\log (\log (N(x)))$ versus log x. Nevertheless,
this procedure may yield imprecise results. Therefore, we propose looking at
the exact ratio $M_{k}=\langle x^{k}\rangle /\langle x\rangle ^{k}$, where $%
\langle x^{k}\rangle $ are the moments of the distribution given by Equation %
\ref{eq:stretched_exp}. For example, when $k=1$, $\langle x\rangle $
corresponds to the average of the distribution given by Equation \ref%
{eq:stretched_exp}. The solution we adopted was to vary $\beta $ so as to
find the best approximation to $M_{k}$ in relation to the experimental
ratios $R_{k}$ calculated by Equation 3. Thus, the values for $M_{k}$ can be
analytically calculated and do not depend on the parameter $x_{0}$:

\begin{equation}
M_{k}=\frac{\Gamma \left( \frac{k+1}{\beta }\right) \Gamma \left( \frac{1}{%
\beta }\right) ^{k-1}}{\Gamma \left( \frac{2}{\beta }\right) ^{k}},
\label{eq:moments}
\end{equation}%
where $\Gamma (z)$ is the gamma function $\Gamma (z)=\int_{0}^{\infty
}t^{z-1}e^{t}dt$.

We then calculate $M_{k}$ for values of $k$ between 1 and 3, using a lag of $%
\Delta k=0.1$, and different values of $\beta =0.20,0.22,...,0.34$ (see
right plot in Figure \ref{Fig:citations_and_ratio}) were considered in the
search for a best match to the experimental ratio $R_{k}$ given by Equation
3.

\begin{equation}
R_{k}=\frac{\sum_{i=1}^{n}x_{i}^{k}}{\left( \sum_{i=1}^{n}x_{i}\right) ^{k}},
\label{eq:experimental_ratio}
\end{equation}%
where $n$ denotes the number of TPC members and is represented by a
continuous curve in the same plot. We can observe that the best match is
found when $\beta =0.28$, corroborating the expected behaviour as described
in \cite{laherrere1998}.

This brief analysis shows that the statistical properties related to the
distribution of the number of citations and its relationship with the \emph{%
h-index} are similar to what was observed in other scientific societies.

Let us also analyse some aspects related to the \emph{h-index} distributions
from TPC members of computer science conferences. A histogram of the \emph{%
h-index} for all collected conferences is illustrated in Figure \ref%
{Fig:Giddings}. Many empirical fits were tested (log-normal, gamma and other
non-symmetric functions). Because of the characteristics of the data, a
normal fit was not attempted. An excellent fit was found by using a function
that comes from Chromatography literature, known as Giddings distribution 
\cite{giddings1954}, defined in the following equation:

\begin{figure}[h]
\centering\includegraphics[width=0.8\textwidth]{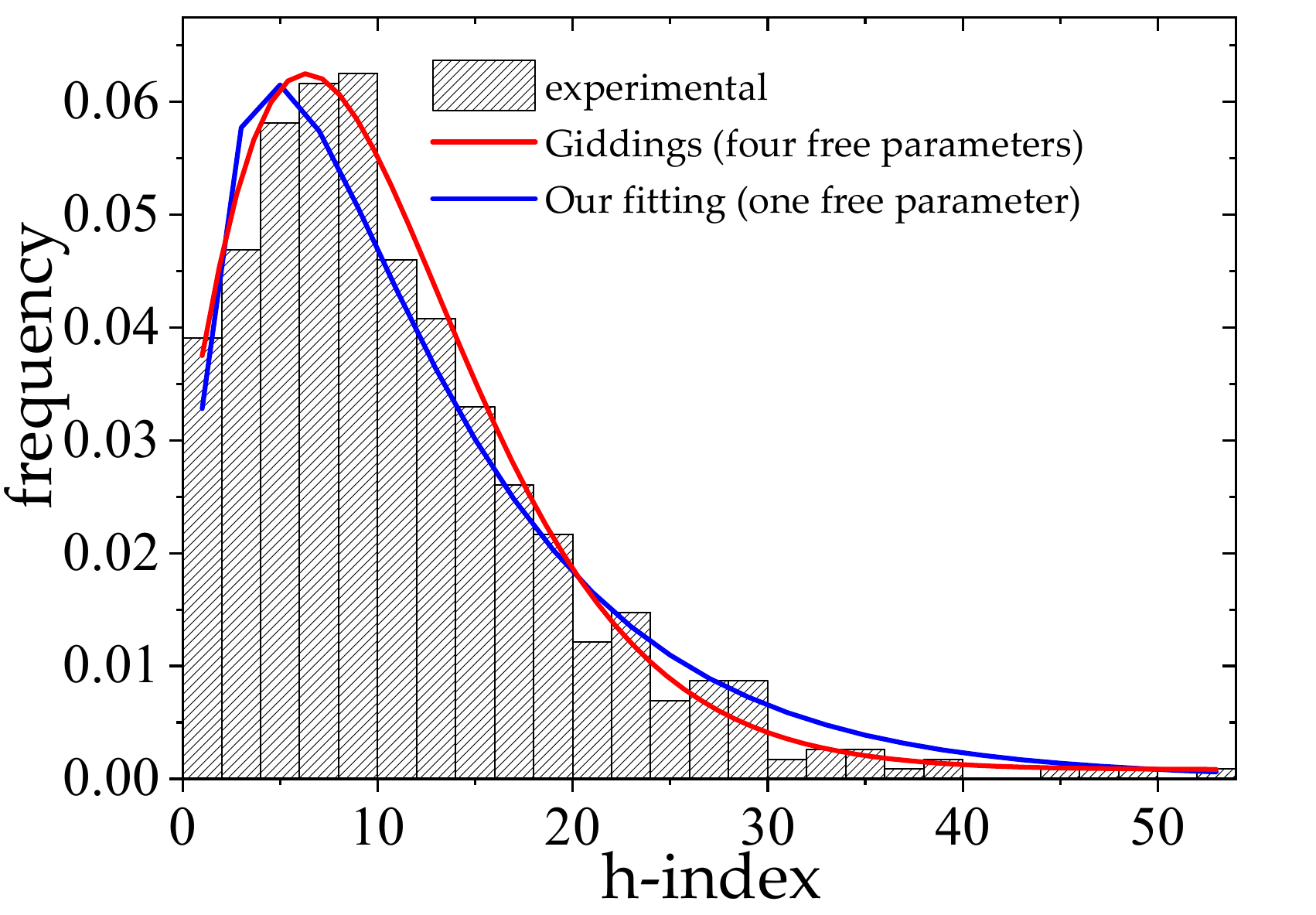}
\caption{Histogram of the \emph{h-index} of TPC members of all conferences.
The best fit (red curve) is the Giddings function (four free parameters). In
blue, we can see our proposed fit with just one parameter.}
\label{Fig:Giddings}
\end{figure}

\begin{equation}
H(h)=H_{0}+\frac{A}{w}\sqrt{\frac{h_{c}}{h}}I_{1}\left( \frac{2\sqrt{h_{c}h}%
}{w}\right) \exp \left( \frac{-h-h_{c}}{w}\right) ,  \label{eq:giddings}
\end{equation}%
where $I_{1}(x)$ is the modified Bessel function, which is described in the
integral form by $I_{1}(x)=\frac{1}{\pi }\int_{0}^{\pi }e^{x\cos \theta
}\cos \theta d\theta $. Apart from the difficulty in analytically evaluating
this function, the fit is numerical and easy to be performed. The fitted
values were $H_{0}=0$, $h_{c}=10.44$, $w=2.518$ and $A=0.91$. The function
given by Equation \ref{eq:giddings} is the distribution in $t$ representing
the chance that one solute molecule will be eluted from the bottom of the
column in a phenomenon of passage of substances through a chromatographic
column (see \cite{giddings1954} for further details). However, fitting a
distribution with four parameters is not simple. A more intuitive formula is
to consider the citation distribution given by Equation (\ref%
{eq:stretched_exp}). By considering Hirsh's relation $x=ah^{2}$, using (\ref%
{eq:stretched_exp}) and after a normalisation, we have that a \emph{h-index}
distribution is given by \cite{daSilva2012}:

\begin{equation}
H_{new}(h) = \frac{2a\beta h}{x_{0}\Gamma(1/\beta)}\exp\left[ - \left( \frac{%
ah^{2}}{x_{0}} \right)^{\beta} \right]  \label{eq:h_distribution}
\end{equation}

Thus, by considering a simple linear fit in a plot of $x$ as function $h^{2}$
we obtain the slope $a$ that for our conferences is $a=5.71$ (using Equation
12). Since we also have $\beta $ previously calculated, varying $x_{0}$
(from $x_{min}=1$ to $x_{max}=90$) we find the best fit, denoted by the blue
curve in Figure \ref{Fig:Giddings}. The best value found for $x_{0}$ by
minimizing the least square function was $x_{0}=51$. $\Gamma $ is the same
gamma function already described in Equation \ref{eq:moments}.

The results of this analysis show a non-symmetric distribution of the \emph{%
h-index} in the program committee of the conferences. But is this indeed a
good feature? In fact, we expect a good conference to have a homogeneous
committee composed of young promising researchers with good \emph{h-indexes}
and also experienced researchers with a good \emph{h-index} achieved through
a sound scientific career. We do not consider a TPC composed of a few
leading scientists padded up with lower-qualified researchers as good. Thus,
in a second investigation, we analyse the \emph{h-index} distribution for
each conference. First, it would be interesting to know if any of the
conferences present a normal distribution of the h-indexes of their TPC
members. Using a traditional Shapiro-Wilk (SW) normality test (see results
in Table \ref{Table:Normality}), we tested the normality level of each
conference studied. The conferences Conf. F and Conf. C (the latter is
normal at a much lower level) were considered to be normally distributed, at
a level of 5\%.

\begin{table}[h]
\caption{Normality analysis of TPC members' \emph{h-indexes}}\centering
\begin{tabular}{lcccc}
\toprule Conference & Normality & $p$-value & Kurtosis & Skewness \\ 
\midrule Conf. D (A) & non-normal & 0.00092 & 0.47117 & 0.70534 \\ 
Conf. A (A) & non-normal & 0.00000 & 9.57315 & 1.99863 \\ 
Conf. F (A) & Normal & 0.45027 & -0.11212 & 0.40324 \\ 
Conf. E (B) & non-normal & 0.00000 & 2.63593 & 1.58998 \\ 
Conf. C (B) & non-normal & 0.00000 & 13.74526 & 3.14864 \\ 
Conf. B (C) & Normal & 0.06612 & -0.24953 & 0.63032 \\ 
Conf. G (C) & non-normal & 0.00007 & 7.99799 & 2.15824 \\ 
\bottomrule &  &  &  & 
\end{tabular}%
\label{Table:Normality}
\end{table}

\begin{figure}[h]
\centering\includegraphics[width=0.5\textwidth]{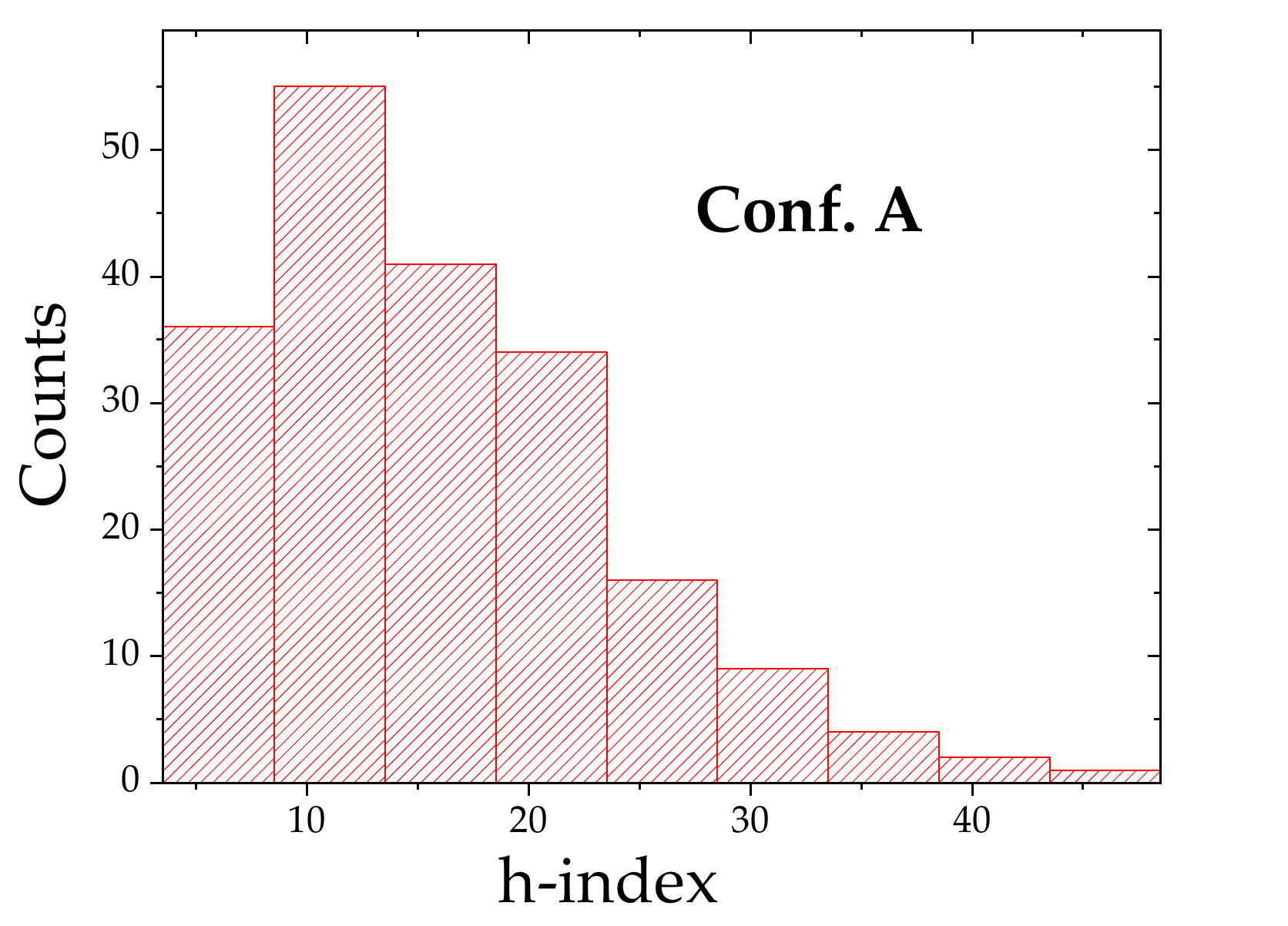}%
\includegraphics[width=0.5\textwidth]{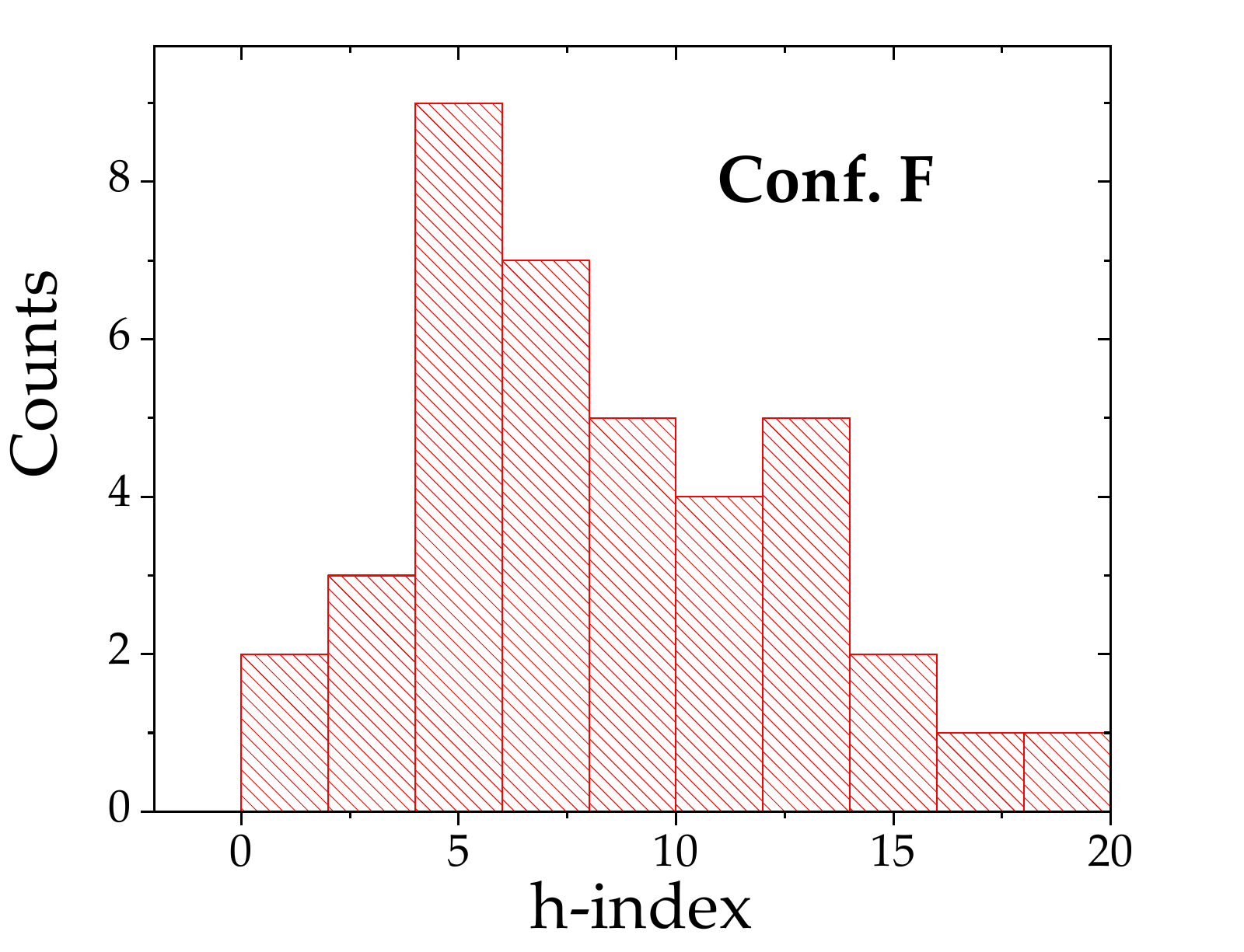}
\caption{\emph{h-index} histograms. The left plot shows a non-normal
conference and the right plot shows a conference that is normal according to
the SW test.}
\label{fig:hists_confA_and_confF}
\end{figure}

Figure \ref{fig:hists_confA_and_confF} shows a graphical comparison between
the \emph{h-index} histograms for Conf. F, which is remarkably normal, and
for Conf. A, which is remarkably non-normal.

For a Gaussian distribution, we expect kurtosis and skewness to approach
zero. We can observe that Conf. A is a conference that does not normal
distribution characterised by a high $p$-value (0.0000), which is more
expected by a natural \emph{h-index} distribution. It presents a heavy tail
(kurtosis = 9.57315) and a good asymmetry (skewness = 1.99863) in relation
to its mean value. As we report in Section 4, Conf. A is the best among the
conferences analysed with our proposed \emph{index}.

There is no conclusive indication regarding the relationship between the
normality of the TPC \emph{h-index} distribution and the quality of the
conference. However, according to our method (presented in the next section)
the best conferences are predominantly non-normal. This suggests good
conferences will likely have many researchers with high \emph{h-indexes} (a
characteristic of heavy tail distribution), despite many of the \emph{%
h-indexes} being concentrated around a mean value. Normality is an
interesting aspect, but it is not the most adequate to determine quality. A
good metric must take into consideration the homogeneity of a group, but, at
the same time, it cannot lose the focus of the magnitude of \emph{h-indexes}
of the members. It is also important to allow comparisons among groups of
different sizes.

Homogeneity together with a reasonable \emph{h-index} definition for groups
(see \cite{egghe2008,schubert2007}) is the main requirement for a good
research group, such as a TPC or an editorial board of a journal. These
aspects are explored in greater detail in the next sections.

\section{The Gini Coefficient and the \emph{h-index} of a group}

The notion of a high-quality group, in any context, presupposes the
excellence of its individual members. In certain cases, however, it is not
sufficient for a group to include merely a few highly productive individuals
alongside others with limited academic output. Homogeneity---understood as a
consistent level of academic productivity among members---is also a
desirable attribute. This is particularly relevant for research groups such
as program committees (TPCs) and editorial boards, where a more homogenous
level of expertise contributes to the fairness and consistency of paper
evaluations.

We deal with that homogeneity enhances the credibility and effectiveness of
such evaluative bodies. Admission to a journal's editorial board or a
conference's TPC should therefore be contingent upon the researcher having
attained a scholarly profile commensurate with the quality standards of the
venue. Given that publication venues vary in their academic rigor,
participation as a reviewer or committee member should not be assumed as a
default entitlement, but rather earned through demonstrated academic
achievement appropriate to the venue's expectations. An interesting
statistic to measure the equality of members in a group comes from the
Social Economics literature, the Gini coefficient \cite{gini1921,gini_wiki}.

In its original formulation, the Gini coefficient (which is a number in the
interval [0, 1]) was designed to quantify inequalities in the distribution
of wealth within a country. The lower the Gini coefficient, the more equal
the wealth distribution. The highest known Gini coefficient is Namibia's
(0.707) while the lowest is Iceland's (0.195) \cite{dorfman1979}. It is
worth mentioning that a low Gini coefficient is positive for a country in
which the population has buying power. Remarkably, countries such as Austria
and Ethiopia have the same Gini coefficient of 0.300. However, this low Gini
coefficient means something good for Austria (a homogeneously high living
standard), but it means something bad for Ethiopia (a homogeneously low
living standard).

To adapt the method for calculating the Gini coefficient to this
bibliometric context, we proceed as follows: first, rank the members of the
population in increasing order of \textquotedblleft
wealth\textquotedblright\ (here represented by the h-index), i.e.,$\
h_{1}<h_{2}<\ldots ,h_{n-1}<h_{n}$. Next, define $\Phi (h_{i})$ as the
fraction of \textquotedblleft bibliometric wealth\textquotedblright\
associated with the fraction of individuals $f_{i}=i/n,i=1,\ldots ,n$, which
is given by:

\begin{equation}
\Phi(h_i) = \frac{\sum_{j=1}^i h_j}{\sum_{j=1}^n h_j}  \label{eq:lorenz}
\end{equation}

\begin{figure}[h]
\centering\includegraphics[width=0.8\textwidth]{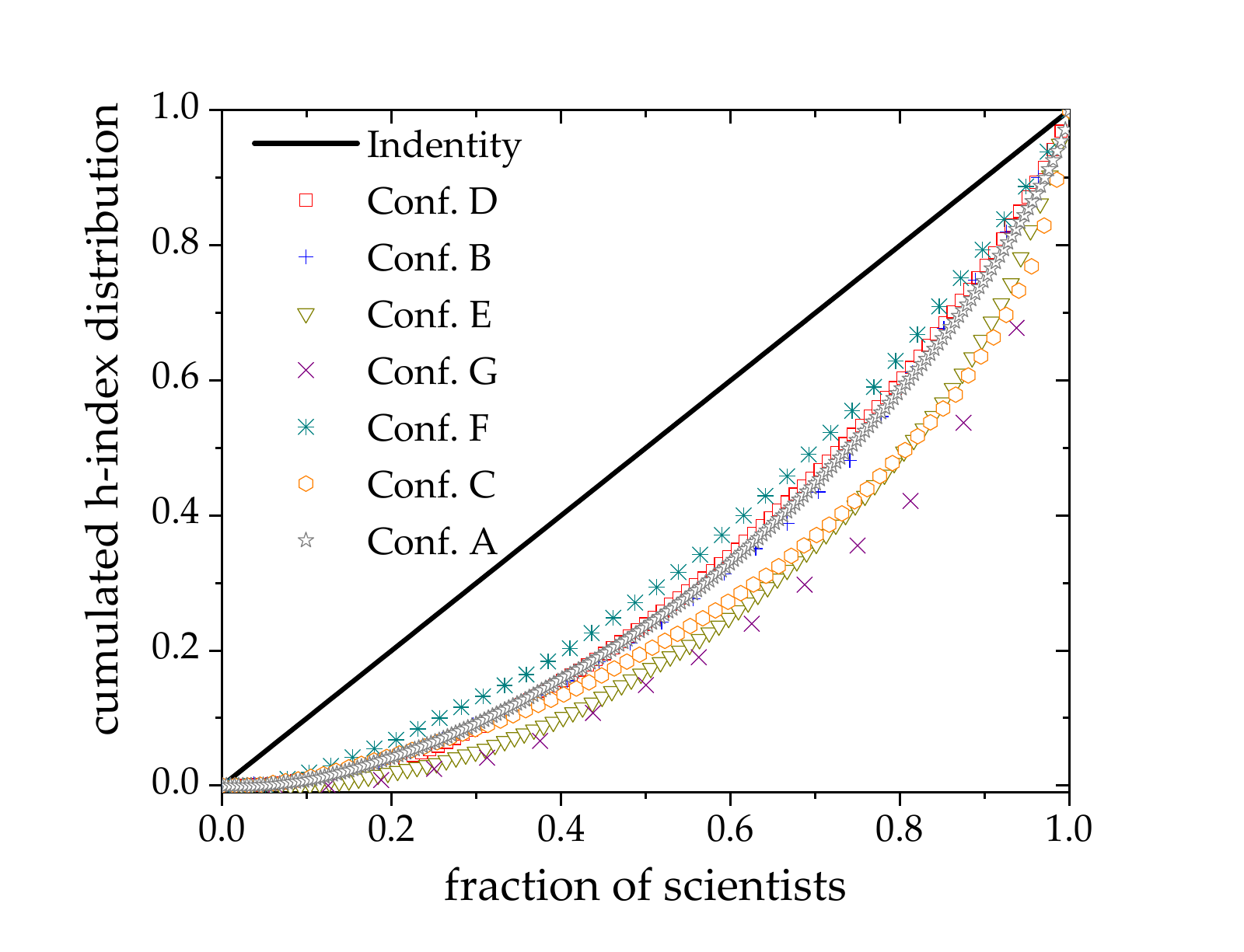}
\caption{Lorenz Curves for the 7 conferences analyzed.}
\label{Fig:Lorenz}
\end{figure}

By applying Eq. \ref{eq:lorenz} to each group, the Lorenz curve \cite%
{gastwirth1972} $(\Phi (h_{i}),f_{i})$ is generated. In a totally fair
society (or TPC), we should expect $\Phi (h_{i})=i/n$, but in real societies
this is not observed. From that, we extend the Lorenz curve concept to
describe inequalities in the \emph{h-index} distribution of the scientific
population, which is presented for the 7 conferences analyzed in this paper
in Figure \ref{Fig:Lorenz}.

We can observe that for each group, the area between the Lorenz curve for
each conference and the perfect \emph{h-index} distribution represented by
the continuous line (identity function $f_i = i/n$) measures the level of
inequality in the conference's TPC. This notion can be quantified by Gini
statistics or simply by the Gini coefficient. The value of the Gini
coefficient is twice the aforementioned area. Theoretically, this
coefficient is calculated as in Equation \ref{eq:gini}.

\begin{equation}
g = 1 - 2\int_{0}^{1}{\Phi(h)dh}  \label{eq:gini}
\end{equation}

Equation \ref{eq:gini} is numerically approximated by a trapezoidal formula,
leading to Equation \ref{eq:gini_numerical}:

\begin{equation}
g=1-\frac{\Phi \left( h_{0}\right) +\Phi \left( h_{n}\right) }{n}-\frac{2}{n}%
\sum_{k=1}^{n-1}{\Phi (h_{k})}=1-\frac{1}{n}\sum_{k=1}^{n}{[\Phi
(h_{k})+\Phi (h_{k-1})]}  \label{eq:gini_numerical}
\end{equation}%
where $\Phi (h_{0})=0$ and $\Phi (h_{n})=1$ for construction.

Our proposed method to classify the quality of a group of researchers from
the \emph{h-indexes} of its members considers the magnitude of the \emph{%
h-index} and the level of equality of this \emph{h-index} in the entire TPC
population. This new definition, which we call "\emph{$\alpha$-index}" is
composed by two different quantities: (i) the Gini coefficient of the \emph{%
h-index} population, and (ii) a definition of the relative \emph{h-index}.

We consider that the \emph{h-index} of a group with $n$ members should be
established by the maximum number of members that have an \emph{h-index}
equal to or higher than an integer $h_{group}$, and necessarily the
remaining ($n-h_{group}$) members have an \emph{h-index} less than $%
h_{group} $.

The groups to be compared may have different numbers of members. Thus, to
compare different groups, we need to define a relative $h_{group}$ which can
be based on the smallest group to be compared. Let us consider the simplest
situation: two groups $r_1$ and $r_2$ with sizes respectively denoted by $%
|r_1|$ and $|r_2|$, with $|r_1| < |r_2|$. Denoting $H^{(2)} = \{h_1^{(2)},
h_2^{(2)}, \ldots, h_{|r_2|}^{(2)}\}$ as the set of \emph{h-indexes} of
members of group $r_2$, we define the relative $h_{group}$ of $r_2$ in
relation to $r_1$, over a number of samples ($n_{sample}$) as the value
calculated by Eq. \ref{eq:h_group}:

\begin{equation}
h_{group}^{(2)}=\frac{1}{n_{sample}}%
\sum_{j=1}^{n_{sample}}h_{group}(S_{j}^{(r_{1})})  \label{eq:h_group}
\end{equation}%
where $S_{j}^{(r_{1})}$ denotes the $j$-th \emph{h-index} sample of size $%
|r_{1}|$ randomly chosen in $H^{(2)}$. This normalisation is required
because the group $r_{2}$ theoretically should have a maximum \emph{h-index} 
$=|r_{2}|$, whereas $r_{1}$ cannot match that value because it has fewer
members. It is important to mention that our definition requires the
gathering of samples of "smallest group size" inside of larger groups in a
way that groups of different sizes can be compared.

\begin{figure}[h]
\centering\includegraphics[width=0.5\textwidth]{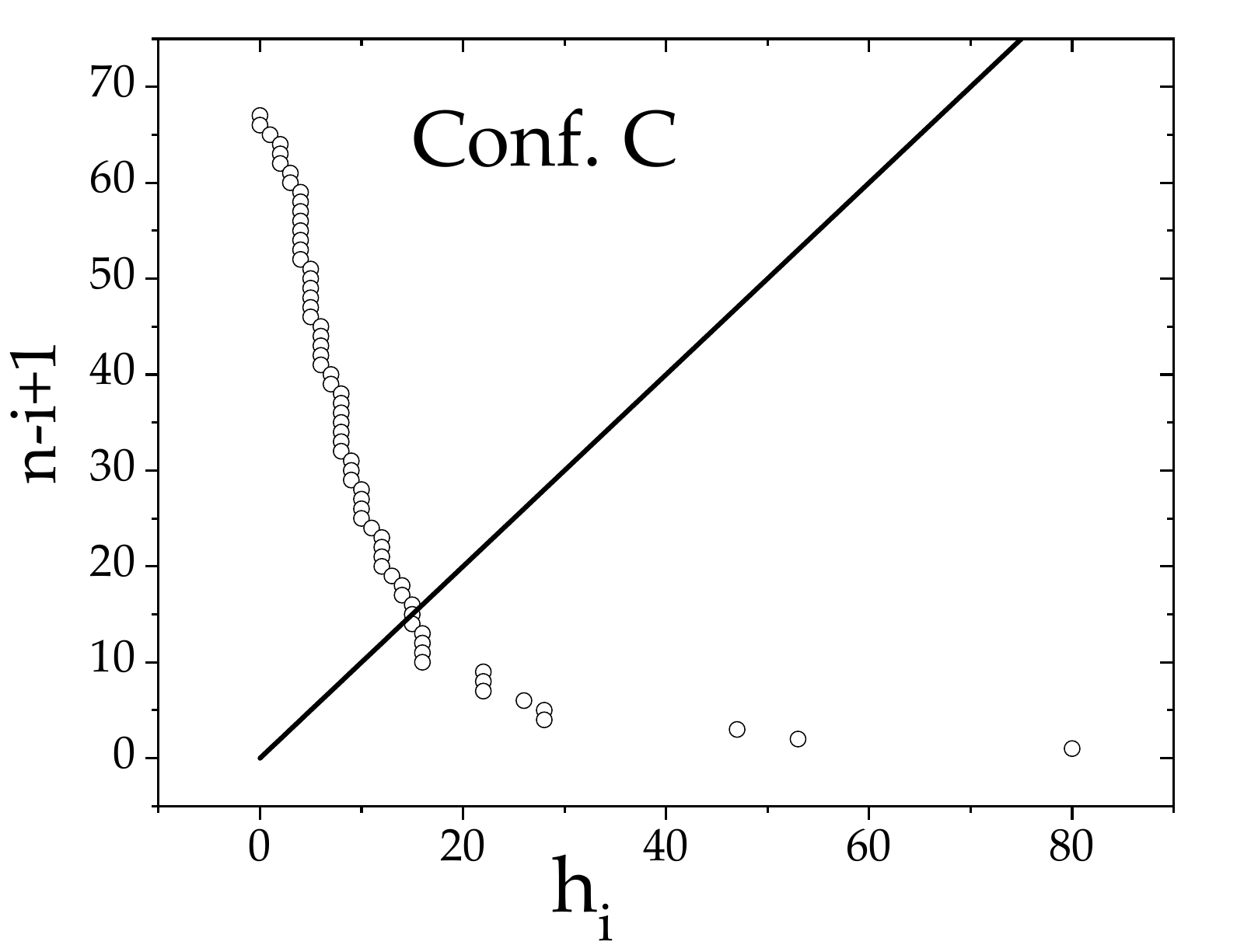}%
\includegraphics[width=0.5\textwidth]{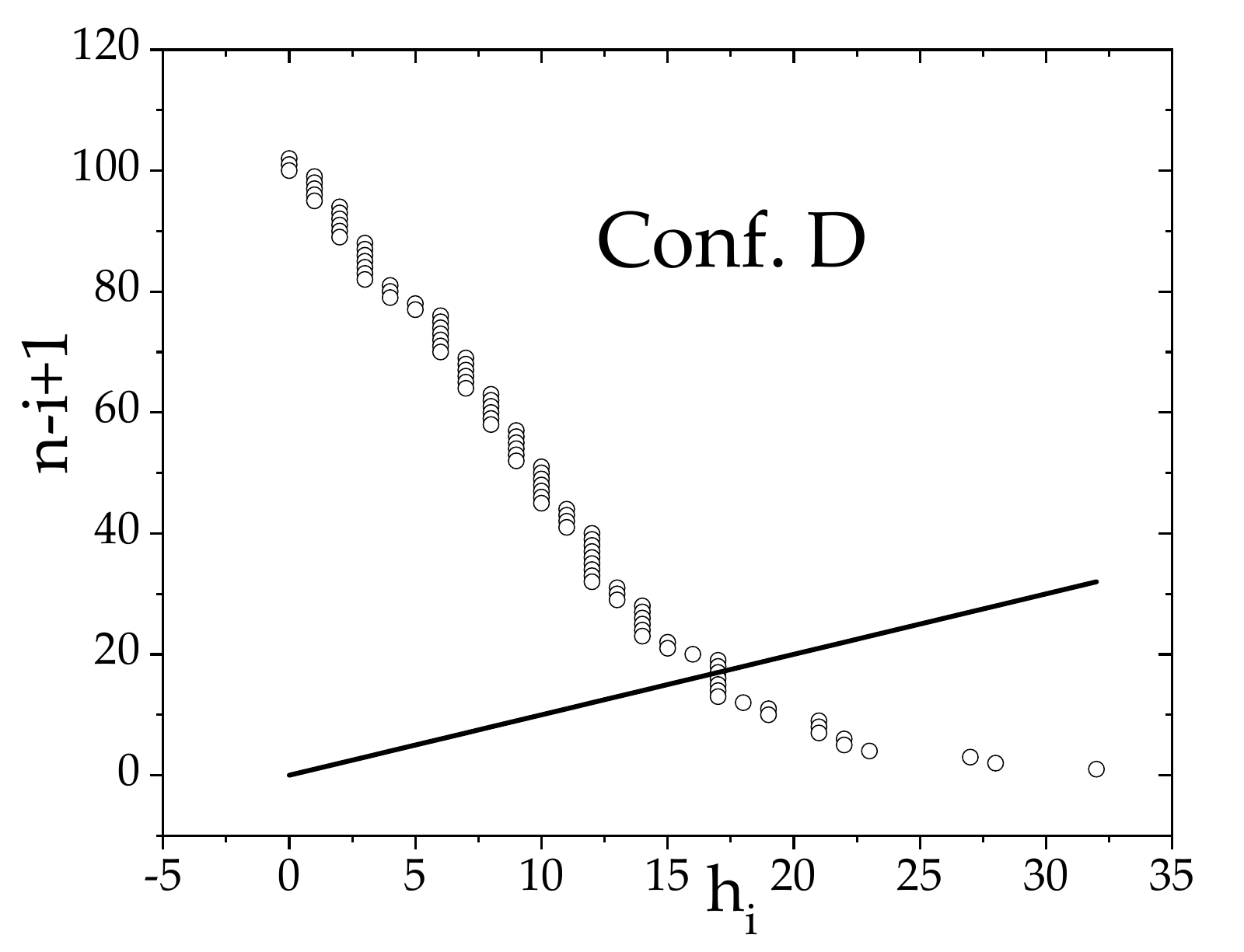}
\caption{Quantity $n-i+1$ as a function of $h_{i}$, $i=1,..,n$ for two
sample conferences. The intercept with the identity line corresponds to the $%
h_{group}$.}
\label{Fig:hgroup}
\end{figure}

In practice, to find the $h_{group}$ it suffices to plot the function $\psi
(h_{i})=n-i+1$ (number of members that have an \emph{h-index} higher than $%
h_{i}$ ) as a function of $h_{i}$ and to determine the intercept between $%
\psi (h_{i})$ and the identity function $\phi (h_{i})=h_{i}$ since $%
h_{i}<h_{i+1}$, for $i=1,\ldots ,n-1$ (see Figure \ref{Fig:hgroup}).

In case of $m>2$ groups, a simple algorithm is proposed:

\begin{enumerate}
\item Input: $m$ groups denoted by $r_1, r_2,\ldots, r_m$, and number of
samples ($n_{sample}$) are required;

\item The smallest group ($r_k$) is identified, i.e., $k = \arg
\min\{|r_l|\}_{l=1}^m$;

\item For each group indexed by $l =1, .., m$, samples $S_j^{(l)}$ of \emph{%
h-index} of size $|r_k|$ are randomly chosen in $H^{(l)} = \{h_1^{(l)},
h_2^{(l)}, \ldots, h_{|r_l|}^{(l)}\}$, with $j = 1, \ldots, n_{sample}$ and
the relative h-group in relation to group $r_k$, i.e., $h_{group}^{(l)}$ is
calculated according to Equation \ref{eq:h_group}.
\end{enumerate}

From that, a ranking for conferences (or groups) can be established based on
their relative $h_{group}$ and the Gini coefficient. Our main proposed
function, the \emph{$\alpha$-index}, is employed to measure the quality of a
group $l$ among $m$ groups. Equation \ref{eq:alpha_index} defines the \emph{$%
\alpha$-index}.

\begin{equation}
\alpha_l = \frac{h_{group}^{(l)}(1 - g_l)}{\max_{i=1}^m \{h_{group}^{(i)}(1
- g_i)\}},  \label{eq:alpha_index}
\end{equation}
where $k = \arg \min\{|r_i|\}_{i=1}^m$ and $g_l$ is the Gini coefficient of
group $l$. The value $0 \leq \alpha_l \leq 1$ measures the quality of a
group based on a convenient definition of the \emph{h-index} for groups
weighed by the Gini coefficient of members in all groups considered for
ranking. The factor $g_l$ works as an amplifier of the relative $h_{group}$.
The smaller $g_l$, the more significant the $h_{group}$.

\section{Experiments}

To evaluate the conferences, the first step is to calculate the relative $%
h_{group}$ based on the size of the smallest program committee among the
conferences (Conf. G, 16 members). For such a calculation, we used the
simple algorithm presented in Section 3. For our computations, we used $%
n_{sample} = 1000$. In Table \ref{Table:hgroup}, we show our proposed ranking for the
conferences according to the \emph{$\alpha$-index}.

Our results show a divergent ranking from the one that would be established
by the simple calculation of the average of the \emph{h-indexes}. Our \emph{$%
\alpha$-index} shows the need for the inclusion of the Gini coefficient or
another homogeneity parameter in the analysis of the quality of conferences.
Many conferences have a high $h_{group}$ due to just a small fraction of the
TPC. The Gini coefficient shows how representative the computed $h_{group}$
is. A low $g$ denotes that the conference has a robust $h_{group}$.
Furthermore, it means that any smaller sample collected from the group
should have the same $h_{group}$, making it independent from the sample.
Conferences with a high Gini coefficient have discrepant TPC members, which
is a sign of questionable quality.

The ranking of conferences according to the \emph{$\alpha$-index} differs
from the ranking by CAPES, the research funding agency, in two cases. Conf.
F is given a lower ranking according to the \emph{$\alpha$-index}. In fact,
in a later assessment, CAPES re-evaluated subsequent editions of this
conference, placing it at a lower rank. The second discrepancy was for Conf.
B, which ranked higher according to the \emph{$\alpha$-index}. CAPES uses a
minimum number of editions as an attribute to determine the quality of the
conferences, and this may have led to the incorrect ranking of Conf. B. Our
approach does not depend on the number of editions as it analyses the
h-index distribution for a specific edition of a conference, and this is one
of its main advantages.

\begin{table}[h]
\caption{Average \emph{h-index}, Gini-Coefficient, $h_{group}$ and $\protect%
\alpha$-index for the 7 conferences analyzed.}\centering
\begin{tabular}{lcccc}
\toprule Conference & avg $h_{index}$ & Gini coef. & $h_{group}$ & $\alpha$%
-index \\ 
\midrule Conf. A (A) & 12.78 $\pm$ 0.65 & 0.377 & 23 & 1.000 \\ 
Conf. D (A) & 10.10 $\pm$ 0.66 & 0.367 & 17 & 0.820 \\ 
Conf. C (B) & 11.63 $\pm$ 1.55 & 0.462 & 15 & 0.700 \\ 
Conf. B (C) & 11.92 $\pm$ 1.60 & 0.381 & 12 & 0.652 \\ 
Conf. E (B) & 08.07 $\pm$ 0.83 & 0.487 & 14 & 0.648 \\ 
Conf. F (A) & 07.94 $\pm$ 0.69 & 0.303 & 10 & 0.570 \\ 
Conf. G (C) & 07.56 $\pm$ 2.39 & 0.548 & 6 & 0.380 \\ 
\bottomrule &  &  &  & 
\end{tabular}%
\label{Table:hgroup}
\end{table}

Another characteristic of the \emph{$\alpha$-index} is that the relative
ordering of the groups would remain the same if only a subset of the
conferences had been compared. Also, if we do pairwise comparisons, for
example between conference X and Y and find that X is better than Y. Then
comparing Y to Z we find that Y is better than Z. By transitivity, a
comparison between X and Z would result in X having a higher \emph{$\alpha$%
-index} than Z.

Finally, other interesting instances for our approach could be easily
experimented. For example, one could consider not only the \emph{h-index} of
the TPC members but also the \emph{h-indexes} of the authors who have
published papers in the conference. The difficulty here is the greater
quantity of data required and its pre-processing.

\section{Summary and Conclusions}

This paper proposed a new method for classifying research groups in any
scientific research area. Our method combines the concepts of homogeneity
(Gini coefficient) and magnitude (relative $h_{group}$) to measure the
quality of a group. Analysing normal and non-normal groups of researchers,
more specifically, program committees of scientific conferences, we
established a ranking for seven conferences. In addition, in a preliminary
analysis, we provided a detailed description of the statistical properties
of the data.

Our results indicate that a fair classification should consider more than
simply a high average \emph{h-index}. Characteristics such as the
homogeneity of the group, evidently with a reasonable \emph{h-index}, should
also be included in the criteria. Our results agree with CAPES's
classification scheme in most cases, but point out some shortcomings in the
agency's classification, showing that a simple implementation of our method
would yield fairer ranking.

Although in this paper we analysed the quality of computer science
conferences and showed how to rank conferences based on the proposed index,
this method can be naturally extended to classify any other group of
researchers in which homogeneity is a desirable feature. The method may be
employed to characterise the quality of a journal by collecting the \emph{%
h-indexes} of the members of its editorial board in a more restricted
database such as ISI-JCR \cite{isi2010}. Our approach could also be used to
establish a comparison between journals and conferences, and maybe between
research areas such as computer science and physics, or even more distant
fields such as the humanities and exact sciences.

\section*{Acknowledgments}

We gratefully acknowledge the partial support provided by CNPq: RS under
grant 304575/2022-4, and JPMO under grant 306695/2022-7.

\bibliographystyle{unsrt}
\bibliography{references.bib}

\end{document}